%% file: template.tex
\documentclass{article}
\usepackage{arxiv}
\usepackage{cite}
\usepackage[utf8]{inputenc} 
\usepackage[T1]{fontenc}    
\usepackage{hyperref}       
\usepackage{url}            
\usepackage{booktabs}       
\usepackage{amsfonts}       
\usepackage{nicefrac}       
\usepackage{microtype}      
\usepackage{lipsum}

\usepackage{graphicx}
\usepackage{amssymb} 
\usepackage{hyperref}

\usepackage[makeroom]{cancel}
\usepackage{algorithm,algpseudocode}
\usepackage{pgf}
\usepackage{tikz}
\usepackage{authblk}
\usepackage{soul} 
\usepackage{mathtools} 
\usepackage{fancyvrb} 
\usepackage{media9}
\usepackage{epigraph}
\usepackage{smartdiagram}
\usepackage{verbatim}
\usepackage{lmodern}
\usepackage{listings}
\usepackage{fancyvrb}
\usepackage{tcolorbox}
\usepackage{wrapfig}
\usepackage{float}
\usepackage{fancybox}

\newenvironment{filecode-0}[1][]
{\minipage{\linewidth}
\lstset{framexleftmargin=7mm,basicstyle=\ttfamily\fontfamily{pcr}\footnotesize,frame=single,numbers=left,#1}}
{\endminipage}

\newenvironment{filecode-1}[1][]
{\begin{minipage}{\linewidth}
\lstset{framexleftmargin=7mm,basicstyle=\ttfamily\fontfamily{pcr}\footnotesize,frame=trBL,numbers=left,#1}}
{\end{minipage}}

\usepackage{subfig}

\usepackage{lipsum}
\usepackage{amsfonts}
\usepackage{graphicx}
\usepackage{epstopdf}

\usepackage{cleveref}

\ifpdf
  \DeclareGraphicsExtensions{.eps,.pdf,.png,.jpg}
\else
  \DeclareGraphicsExtensions{.eps}
\fi

\usepackage{amsopn}

\title{Performance Optimizations of Recursive Electronic Structure Solvers targeting Multi-Core Architectures\\
LA-UR-20-26665}

\author{
Adetokunbo A. Adedoyin,
Los Alamos National Laboratory, CCS-7 Division,
\texttt{aadedoyin@lanl.gov}
\and 
Christian F. A. Negre,
Los Alamos National Laboratory, T-1 Division,
\and 
Jamaludin Mohd-Yusof, Los Alamos National Laboratory, CCS-7 Division
\and
Nicolas Bock, Los Alamos National Laboratory, CCS-3 Division
\and
Daniel Osei-Kuffuor, Lawrence Livermore National Laboratory, CASC Division
\and
Jean-Luc Fattebert, Oak Ridge National Laboratory, CSE Division
\and
Michael E. Wall, Los Alamos National Laboratory, CCS-3 Division
\and
Anders M. N. Niklasson, Los Alamos National Laboratory, T-1 Division
\and
Susan M. Mniszewski, Los Alamos National Laboratory, CCS-3 Division, 
\texttt{smm@lanl.gov}

}

\begin{document}
\maketitle

\begin{abstract}
\input{abstract}

\end{abstract}


\section{Introduction}
\input{introduction}
\input{methodology}

\input{results}
\input{summary}

\section*{Acknowledgments}
We would like to acknowledge the assistance of volunteers in putting
together this example manuscript and supplement. This work was performed as part of the Co-design Center for Particle Applications, 
supported by the Exascale Computing Project (17-SC-20-SC), a collaborative effort
of the U.S. DOE Office of Science and the NNSA. This work was performed under the U.S. Government contract 89233218CNA000001 for
Los Alamos National Laboratory (LANL), U.S. Government Contract DE-AC52-07NA27344
for Lawrence Livermore National Laboratory (LLNL), U.S. Government Contract
DE-AC05-00OR22725 for Oak Ridge National Laboratory (ORNL)

\bibliographystyle{unsrt}  

\bibliography{references}

\end{document}

%% file: abstract.tex
\section{Abstract}
As we rapidly approach the frontiers of ultra large computing resources, software optimization is becoming of 
paramount interest to scientific application developers interested in efficiently leveraging all available 
on-Node computing capabilities and thereby improving a requisite science per watt metric. The scientific 
application of interest here is the Basic Math Library (BML) that provides a singular interface for linear 
algebra operation frequently used in the Quantum Molecular Dynamics (QMD) community. The provisioning of a 
singular interface indicates the presence of an abstraction layer which in-turn suggests commonalities in 
the code-base and therefore any optimization or tuning introduced in the core of code-base has the ability to 
positively affect the performance of the aforementioned library as a whole. With that in mind, we proceed 
with this investigation by performing a survey of the entirety of the BML code-base, and extract, in form 
of micro-kernels, common snippets of code. Since the data structure of the core BML code-base is well 
established we pursue less invasive optimization strategies, that is, we focus our effort on optimizing at 
the thread level as opposed to modifying the data-structures for performance at the Single Instruction Multiple Data (SIMD) scale. 
We incorporate several active and passive directives/pragmas that aid to inform the compiler on nature of the 
data-structures and algorithms. Following that, we introduce several optimization strategies into 
these micro-kernels including 1.) Strength Reduction 2.) Memory Alignment for large arrays 3.) Non Uniform Memory Access (NUMA) 
aware allocations to enforce data locality and 4.) appropriate thread affinity and bindings to enhance 
the overall multi-threaded performance. After introducing these optimizations, we benchmark the micro-kernels and compare 
the run-time before and after optimization for several target architectures. Finally we use the results as 
a guide to propagating the optimization strategies into the BML code-base. As a demonstration, herein, we test 
the efficacy of these optimization strategies by comparing the benchmark and optimized versions of the code 
using a 1.) matrix-matrix multiplication using the ELLPACK format and 2.) a full simulation using ExaSP2, 
a proxy application for linear scaling electronic structure calculations. The results of the optimization are 
promising and in agreement with findings in literature.

%% file: introduction.tex
\section{Introduction}
The optimization of electronic structure codes requires different efforts including the improvement of 
solvers, and the adaptation of basic operations to novel computer architectures. With the 
advent of the exascale computing architectures, the aforementioned adaptation requires thorough modifications 
in order to maximize the use of the computational capacity \cite{noauthor_undated-za,CoPA-Exascale}. Furthermore, in the specific case of electronic structure calculations, different chemical systems require specific solver as well as specific architecture adaptation, and generally, what is beneficial for some systems will not necessarily be for others.
QMD technique requires solving for the electronic structure of the physical system to advance the positions of the atoms at each simulation time-step. Solving for the electronic structure is generally a task that involves a high computational cost due to the fact that the number of arithmetic operations scale with the cube of the number of atoms. In order to alleviate this computational cost, various $\mathcal{O}(N)$ complexity algorithms have been proposed \cite{Bowler2012-ew,Goedecker1999-wt}. These algorithms typically approximate the solution with an error that can be controlled by means of an adjustable parameter. They are typically iterative, and require linear algebra operations in each iteration \cite{Niklasson2002-rl,Mniszewski2019-qt}. For optimal performance however, these linear scaling codes have to be optimized at both the vector and thread level. 

Optimizing data structure for improved parallelism is challenging and can be somewhat disruptive, therefore, we focus on optimizing at the thread level.
%
%
%
One of the primary purposes of multi-core computer technology is latency hiding. However, at the cost of latency
hiding is the inherent programmability challenges needed to be circumvented in order to achieve high performance 
gains in software applications, measured in terms of improvements in floating point operations (FLOPS) or run-time. Therefore, herein, we attempt 
to optimize the Basic Matrix Library (BML) software for performance gains in terms of run-time. 
BML is a collection of matrix data formats (for dense and sparse) and basic matrix operations designed to help electronic structures codes run efficiently on various platforms~\cite{Bock2018}\cite{bml_github}\cite{bml}.
We focus on several strategies namely 
(i) Strength Reduction (SR), (ii) memory alignment (MA) to prevent cache contention, (iii) memory initialization with an 
understanding of first touch (FT) policy on Linux system, and (iv) thread affinity and binding (A\&B) optimizations that 
works in conjunction with all the aforementioned techniques for optimal performance. 

The rest of this manuscript is organized in the format delineated below. Section~\ref{section:methodology} expands 
on the optimizations techniques implemented in the BML~\cite{bml} software. It is preceded by section~\ref{section:brief}, 
where we brief on the overall approach taken to optimizing the BML~\cite{bml} software. This is followed by section~\ref{section:targetarchitecture}, which discusses the target computer architectures used for evaluating 
the performance of the BML~\cite{bml} code-base before and after optimization. 
\cref{section:SR,section:NUMA-allocations,section:MA,section:AB} discusses and demonstrates in detail the 
optimization techniques introduced into the BML~\cite{bml} software namely Strength Reduction (SR)~\cite{park2012efficient,godbolt2020optimizations,adedoyin2017,vladimirov2015fine,
waite2012compiler,sheldon2001strength,amarasinghe1999strength,cocke1977algorithm,allen1981reduction}, NUMA aware allocations to enforce data locality~\cite{tate2014programming,denoyellenuma,adedoyin2017,Asai2017-1,Asai2017-2}, Memory Alignment (MA)~\cite{Intel-2013-1,ALESSANDRINI2016375,JEFFERS2013107,
Eltablawy2017,Vladimirov2015,Application2017,Asai2017-5,Vladimirov2012} for large arrays and appropriate thread affinity and bindings (AB)~\cite{VictorEijkhout-1,eijkhout2017parallel,Intel-2019-1,Intel-2019-2,openmp-1}
to enhance multi-threaded performance, respectively. \cref{section:SR,section:NUMA-allocations,section:MA,section:AB} 
is also accompanied by sample implementation of in form of code snippets demonstrating the implementation of the 
aforementioned optimizations techniques. It also shows individual results of performance improvements. 
\cref{section:results-1} combines all the aforementioned optimization techniques and introduces them into the 
BML software \cite{bml}. Following that we generate several pseudo system matrices representing metals, semi conductors and 
soft matter ranging in size from 1000 to 32000 and evaluate the efficacy  of these optimizations by measuring the 
performance differences of BML's EllPACK matrix-matrix multiply algorithm before and after tuning \cite{bml}. \cref{section:results-2} 
evaluates the performance of ExaSP2, a proxy application for performing QMD calculations that relies on the BML~\cite{bml} 
software for its linear algebra computations. System matrices representing semi conductors and soft matter of size 32000 are evaluated for performance. \cref{section:conclusion} contains a discussion summary of the findings herein. 


%% file: methodology.tex
\section{Methodology}
\label{section:methodology}
\subsection{Brief}
\label{section:brief}
Performance optimization of a large code-base can be a daunting task therefore, herein we introduce carefully 
designed subroutines, henceforth referred to as micro-kernels, that are representative of the methods in the 
un-optimized BML~\cite{bml} software. For the ease of evaluating the effects of the optimizing techniques introduced herein, 
we design the micro-kernels such that complex code behaviour e.g. cache trashing, branching and non-unit stride 
access are avoided though present in BML's~\cite{bml} ELLPACK subroutine. At a latter time, we plan to address the effects
of such complex behaviour as they impede on overall performance. We proceed by optimizing this micro-kernels in 
a step by step manner while being mindful of the compile time difference that may be introduced while building 
and linking the actual BML software \cite{bml}. The target algorithm is memory bandwidth bound as it is a sparse matrix-matrix 
multiply algorithm with the ratio of FLOP to byte of data (n) requested from memory of $\textup{O}(\textup{n}<3)$ .

\subsection{Target Architecture:}
\label{section:targetarchitecture}
The target multi-core platform experimented with herein are recent releases of Intel architectures namely Intel's 
Sky-lake Gold, Sky-lake Platinum, Cascade Lake with and without support for Intel's Optane\texttrademark DC persistent 
memory. Table~\ref{tab:hardware__specs} contains the specification details of the aforementioned architectures. 
%
\input{tables/arch}
%

\subsection{Micro-Kernel Performance Evaluation - Strength Reduction:}
\label{section:SR}
\begin{figure}[h!]
    \includegraphics[width=\textwidth]{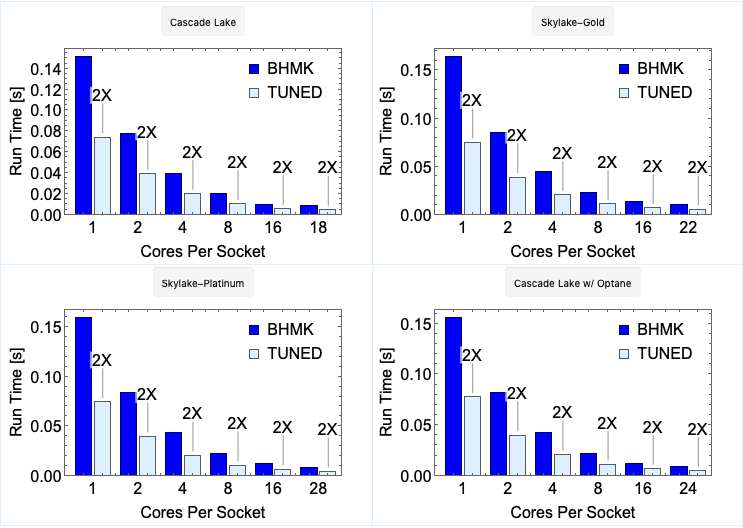}
    \caption{Performance in run-time for the benchmark (BHMK) vs. optimized (TUNED) micro-kernel for SR
    for Cascade Lake (Top-Left), Sky-Lake Gold (Top-Right), Sky-Lake Platinum (Bottom-Left) and 
    Cascade Lake with Intel Optane Technology (Bottom-Right).}
    \label{fig:SR-All}
\end{figure}

Strength Reduction~\cite{park2012efficient,godbolt2020optimizations,adedoyin2017,vladimirov2015fine,
waite2012compiler,sheldon2001strength,amarasinghe1999strength,cocke1977algorithm,allen1981reduction}, is an
optimization procedure, driven either via human intervention or software compiler, where by high latency (costly) 
operations are substituted with their lower latency (cheaper) counterpart while maintaining mathematical 
correctness. SR or approach approximate strength reduction (ASR) techniques can vary from very simple 
~\cite{adedoyin2017,vladimirov2015fine} to more complex and involving substitutions\cite{park2012efficient,
godbolt2020optimizations,waite2012compiler,sheldon2001strength,amarasinghe1999strength,cocke1977algorithm,
allen1981reduction}. The need for SR tuning in the BML~\cite{bml} software, though not prevalent, is a low hanging fruit 
therefore we evaluate the performance difference between atypical occurrences of in-loop division with 
multiplications. Listing~\ref{lis:strengthreduction-OFF} is a code snippet of the micro-kernel atypical of the 
occurrences in the BML software \cite{bml}. Listing~\ref{lis:strengthreduction-ON} is a SR substitutions where we replace 
divisions within a loop with a single multiplication. Figure~\ref{fig:SR-All} is a run-time comparison of the 
benchmark (BHMK) vs. optimized (TUNED) micro-kernels for multiple dual socket Intel architectures using Intel 17
compiler. The performance of the optimized (TUNED) micro-kernel is approximately 2X across multiple thread sizes
and varying architectures. 
\begin{figure}
\centering
\begin{filecode-0}[]
\begin{lstlisting}
#pragma omp parallel for simd
#pragma vector aligned
for(i=0; i < A->N; i++)
{
 A->index[i] = i
 A->nnn[i]  += i
 A->value[i] = 16.0/RAND_MAX;
}
\end{lstlisting}
\end{filecode-0}
\caption{
Micro-kernel without strength reduction representative of the subroutines in BML~\cite{bml} before tuning.}
\label{lis:strengthreduction-OFF}
\end{figure}
\begin{figure}
\begin{filecode-1}[]
\begin{lstlisting}
double  INV_RAND_MAX = 1.0/RAND_MAX;
#pragma omp parallel for simd
#pragma vector aligned
for(i=0; i < A->N; i++)
{
 A->index[i]  = i
 A->nnn[i]   += i
 A->value[i]  = 16.0*INV_RAND_MAX;
}
\end{lstlisting}
\end{filecode-1}
\caption{Micro-kernel with strength reduction applied.}
\label{lis:strengthreduction-ON}
\end{figure}
%
\subsection{Micro-Kernel Performance Evaluation - NUMA Aware Data}
\label{section:NUMA-allocations}
\begin{figure}[h]
\includegraphics[width=\textwidth]{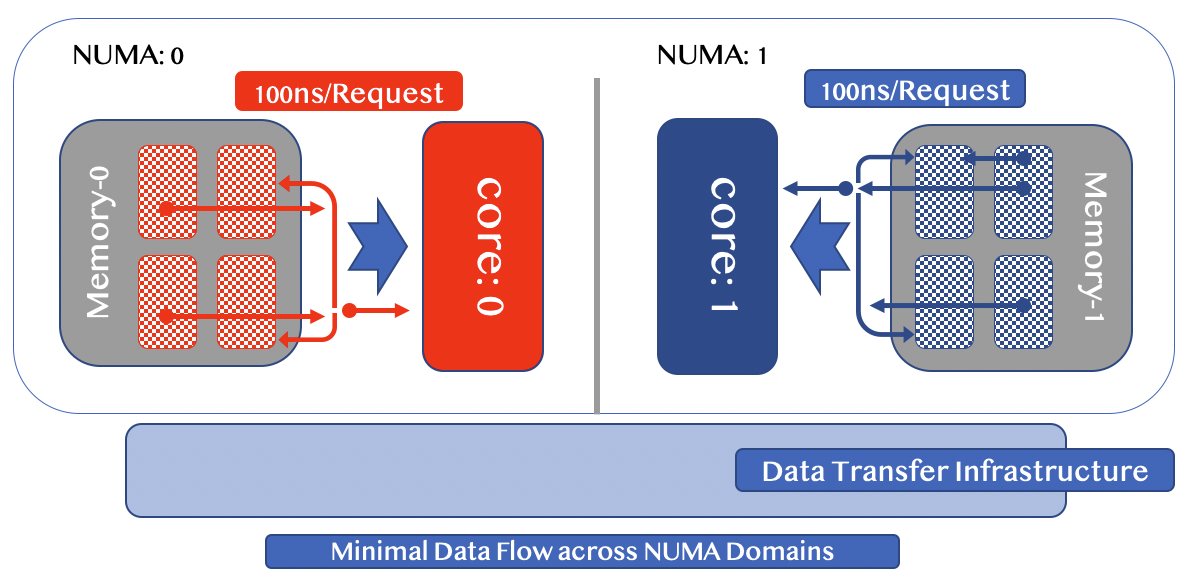}
\caption{Illustration of modern multi-core architectures with multiple NUMA domains. Showing the effect of 
programming with considerations for ``First Touch” policies on Linux systems. During simulation data present 
on NUMA:0 is fetched from NUMA:0 and vice-versa.}
\label{fig:NUMA__aware}
\end{figure}

\begin{figure}[h!]
    \includegraphics[width=\textwidth]{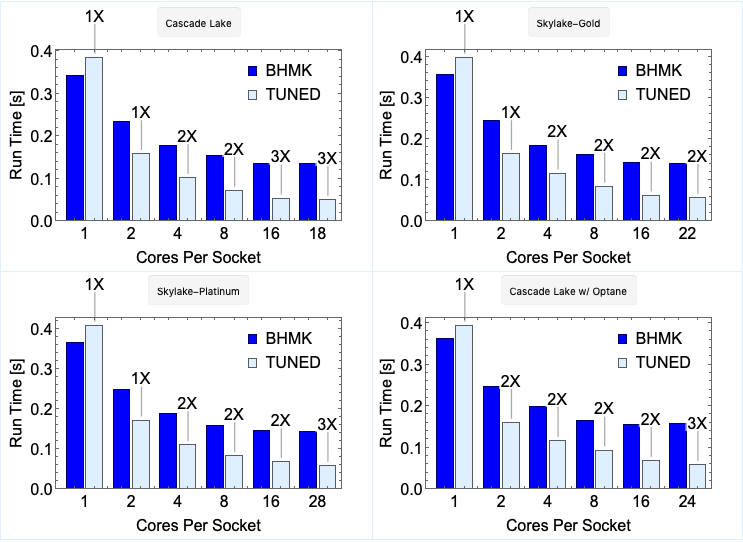}
    \caption{Performance in run-time for the benchmark (BHMK) vs. optimized (TUNED) micro-kernel 
    for data locality for Cascade Lake (Top-Left), Sky-Lake Gold (Top-Right), Sky-Lake Platinum 
    (Bottom-Left) and Cascade Lake with Intel Optane Technology (Bottom-Right).}
    \label{fig:FT-All}
\end{figure}

Data locality in the BML~\cite{bml} software is tuned for by ensuring that 1.) malloc-ed memory is initialized 
in a parallel region with the ``First Touch'' policy in mind and 2.) the affinity and binding settings for 
the initialization and computation loops for data used in multi-threaded regions are specified carefully such that computational data 
remains NUMA local. The ``First Touch'' policy with respect to memory allocation and memory page assignment 
on Linux systems, is associated with the physical location of memory (NUMA~\cite{lameter2013numa} domain) at the point in which 
the actual memory addresses get modified (initialization). At the point of modification, the memory pages get 
assigned and further associated with that specific NUMA domain. Therefore, the First Touch policy determines 
memory page ownership \cite{Intel-2013-1}. Consideration for NUMA in order to avoid needless data movement (data locality) can 
be implemented to cater to different levels of parallelism which may include the SIMD, Thread and Node level. 
Several researchers~\cite{tate2014programming,denoyellenuma,adedoyin2017,Asai2017-1,Asai2017-2} have shown that the 
aforementioned approach improves the overall performance of computational algorithms by avoiding needless data 
movement. Prior to optimization for data locality, atypical linear algebra operations performed using the BML API 
required three steps namely 1.) memory allocation and initialization followed by 2.) the linear algebra calculation
of interest and 3.) Memory de-allocation. The memory allocation and initialization step above is typically achieved 
in BML via two calls, one to \verb|malloc()| and the other to \verb|memset()|. This is also achieved via a single call to \verb|calloc()|.
\begin{figure}
\begin{filecode-0}
\begin{lstlisting}
// Data Allocation
A->nnz   = (int *)    malloc(sizeof(int)*N);
A->index = (int *)    malloc(sizeof(int)*N*M);
A->value = (double *) malloc(sizeof(double)*N*M);

// Non-NUMA Aware: Initialization
memset(A->nnz,   0,   A->N*sizeof(int));
memset(A->index, 0,   A->N*A->M*sizeof(int));
memset(A->value, 0.0, A->N*A->M*sizeof(double));

// Computational Loop
#pragma omp parallel for simd
#pragma vector aligned
for(i=0; i < A->N; i++)
    compute(A);

\end{lstlisting}
\end{filecode-0}
\caption{Micro-kernel without consideration for data locality representative of the subroutines in BML~\cite{bml} before tuning.}
\label{lis:nonnumaaware}
\end{figure}
\begin{figure}
\begin{filecode-1}
\begin{lstlisting}
// Data Allocation
A->nnz   = (int *)    malloc(sizeof(int)*N);
A->index = (int *)    malloc(sizeof(int)*N*M);
A->value = (double *) malloc(sizeof(double)*N*M);

// NUMA Aware: Initialization:
#pragma omp parallel for simd
#pragma vector aligned
    for(i=0; i < A->N*A->M; i++)
        initialize(A);
      
// Computational Loop       
#pragma omp parallel for simd
#pragma vector aligned
for(i=0; i < A->N; i++)
    compute(A);
\end{lstlisting}
\end{filecode-1}
\caption{Micro-kernel with consideration for data locality representative of the subroutines in BML~\cite{bml} after tuning.}
\label{lis:numaaware}
\end{figure}
The second step involving the Linear algebra operation of interest, e.g. a Matrix-Matrix multiply calculation, is
typically comprised of a multi-threaded loop in a parallel region written using the OpenMP paradigm. With the First 
Touch policy in mind, step one and two are in conflict given that they are performed in a serial and parallel 
region, respectively. The first step is a serial operation with an associated performance penalty as the memory 
pages become associated with the specific NUMA domain they were first touched. Listing~\ref{lis:nonnumaaware} is a 
code snippet of the micro-kernel atypical of the occurrences in the BML software. Listing~\ref{lis:numaaware} is a 
NUMA aware data initialization that ensures data locality in the computational loop by first touching data in a 
parallel region (lines 7-10) a opposed to (listing~\ref{lis:nonnumaaware} lines 7-9) using \verb|memset()| to initially 
modify malloc-ed data. Figure~\ref{fig:NUMA__unaware}
is a pictorial illustration of a simplified modern multi-core computer showing computational cores and memories 
(RAM) associated with two NUMA domains. NUMA domain zero (core:0 and Memory:0) and one (core:1 and Memory:1) have 
an associate compute core and memory colored in blue and red respectively. To minimizes obscurity, we simplified the 
physical description of each NUMA domain (0 and 1) and further assume that each core is actually multiple cores 
each with multiple levels of associated cache hierarchy and bandwidth infrastructure. The blue banks of memory 
located on NUMA:0 (red) represent data initialized on Memory:0 that potentially get moved to NUMA:1 (blue) during 
computation. This movement leads to additional performance penalty (higher latency) when performing computation on 
non-local data. This needles computational expense may also lead to cross NUMA domain cache conflict. 
Figure~\ref{fig:NUMA__aware} shows a snap-shot of the preferred association of data and computational core in a 
compute loop. A completely disjoint subset of data associated with distinct NUMA domains is ideal, though 
in practice may require careful software design. Figure~\ref{fig:FT-All} is run-time comparison of the 
benchmark (BHMK) vs. optimized (TUNED) micro-kernels for multiple dual socket Intel architectures using Intel 17
compiler. The performance of the optimized (TUNED) micro-kernel is at best 3X for high thread counts and 2X on 
average on all architectures. In conjunction with NUMA aware allocations is the careful specification of thread 
affinity and binding to ensure thread placement is consistent with data location, in addition, the 
specified number of threads per core is consistent with the arithmetic intensity of the application under 
consideration, and finally that threads migration is avoided. Others authors~\cite{9101899,hildenbrand2020dynamic,
mason2020unexpected,barrera2020modeling,wagle2018numa,wang2016predicting} have demonstrated the challenges 
associated with non-locality of data and have proposed several solutions.
\begin{figure}[h!]
\includegraphics[width=\textwidth]{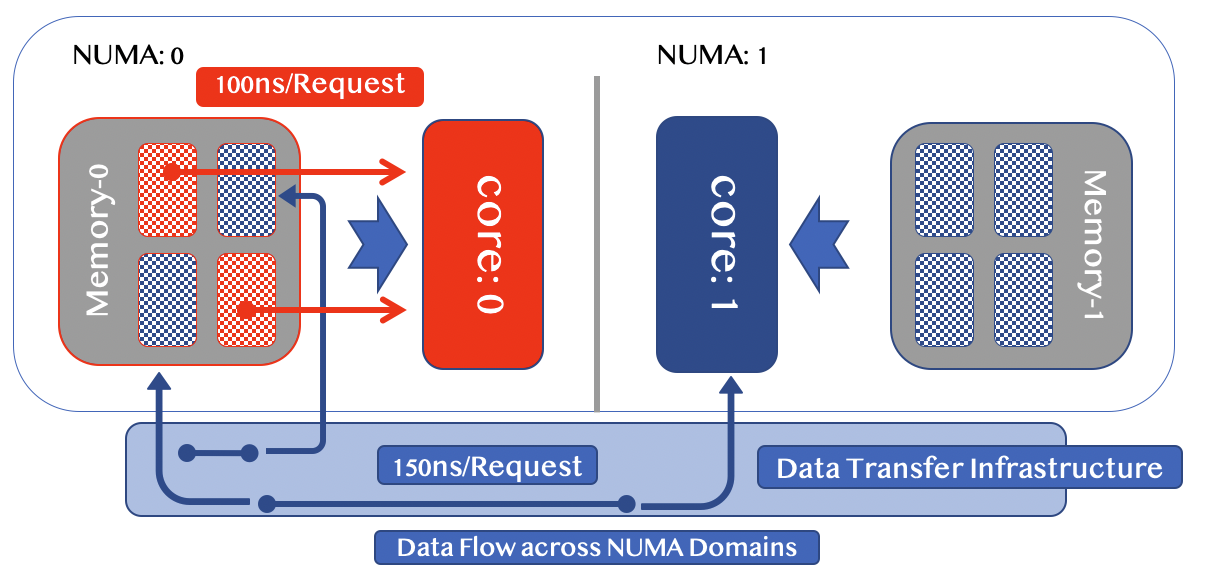}
\caption{Illustration of modern multi-core architectures with multiple NUMA domains. Showing the effect of 
programming without considerations for ``First Touch” policies on Linux systems. During simulation data present 
on NUMA:0 is fetched from NUMA:1. }
\label{fig:NUMA__unaware}
\end{figure}
%
\subsection{Micro-Kernel Performance Evaluation - Memory Alignment:}
\label{section:MA}
\begin{figure}
\centering
\includegraphics[scale=0.5,angle=90]{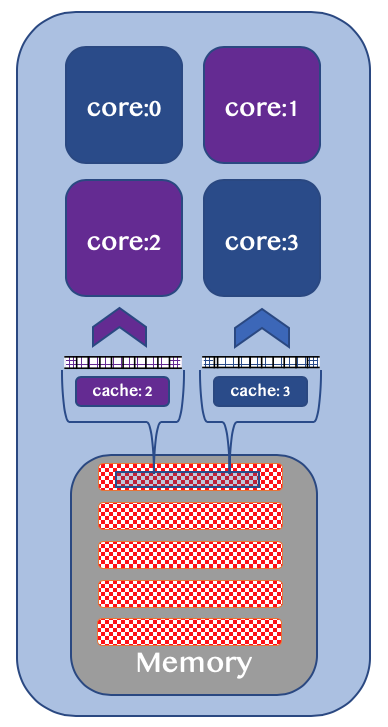}
\caption{An Illustration of the need for memory alignment on multi-core architectures. Showing contention 
between core:2 and core:3.}
\label{fig:MA-Illustration}
\end{figure}

\begin{figure}
\includegraphics[width=\textwidth]{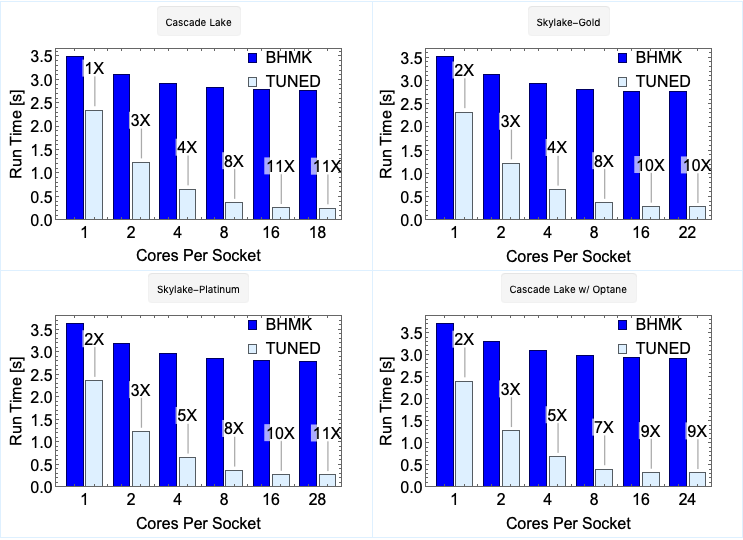}
\caption{Performance in run-time for the benchmark (BHMK) vs. optimized (TUNED) micro-kernel 
         for memory alignment for Cascade Lake (Top-Left), Sky-Lake Gold (Top-Right), 
         Sky-Lake Platinum (Bottom-Left) and Cascade Lake with Intel Optane Technology (Bottom-Right).}
\label{fig:MA-All}
\end{figure}

The application of memory alignment, as demonstrated by~\cite{Intel-2013-1,ALESSANDRINI2016375,JEFFERS2013107,
Eltablawy2017,Vladimirov2015,Application2017,Asai2017-5,Vladimirov2012}, is primarily for the purpose of 
preventing cache contention on multi-core computer hardware. Cache contention on multi-core hardware 
occurs when multiple hardware threads attempt to use the same line of cache. Figure~\ref{fig:MA-Illustration} 
is an illustration of a simplified modern multi-core computer architecture containing four cores and showing 
two cache lines belonging to core:2 (purple) and core:3 (blue). Listing~\ref{lis:MA-OFF} is a code snippet of 
the micro-kernel atypical of the occurrences of memory allocation in the BML software. Listing~\ref{lis:MA-ON} 
is a code snippet of the micro-kernel reflecting memory alignment that ensures minimal cache contention in the 
computational loop. Listing~\ref{lis:MA-ON} shows aligned memory allocations using Intel compiler
~\cite{Intel-2013-1} (lines 2-4), hinting the compiler (lines 11-13) and a complementary aligned memory 
de-allocation (lines 18-20). A similar API that allows for aligned memory allocation is available with the GCC 
compiler~\cite{Linux-1}.
\begin{figure}
\begin{filecode-0}
\begin{lstlisting}
// Allocating unaligned memory:
A->nnz   = (int *)    malloc(sizeof(int)*N);
A->index = (int *)    malloc(sizeof(int)*N*M);
A->value = (double *) malloc(sizeof(double)*N*M);

// Computational loop:
#pragma omp parallel for 
    for(i=0; i < A->N; i++)
        compute(A);

// De-allocation of unaligned memory:
free(A->nnz);
free(A->index);
free(A->value);
\end{lstlisting}
\end{filecode-0}
\caption{Memory allocation in BML software.}
\label{lis:MA-OFF}
\end{figure}
\begin{figure}
\begin{filecode-1}
\begin{lstlisting}
// Allocating aligned memory:
A->nnz   = (int *)    _mm_malloc(sizeof(int)*A->N,64);
A->index = (int *)    _mm_malloc(sizeof(int)*A->N * A->M,64);
A->value = (double *) _mm_malloc(sizeof(double)*A->N*A->M,64);

// Computational loop:
#pragma omp parallel for simd
#pragma vector aligned
for(i=0; i < A->N; i++)
{
 __assume_aligned(A->nnz,64);
 __assume_aligned(A->index,64);
 __assume_aligned(A->value,64);
    compute(A);
}

// De-allocation of aligned memory:
_mm_free(A->nnz);
_mm_free(A->index);
_mm_free(A->value);
\end{lstlisting}
\end{filecode-1}
\caption{Memory allocation with memory alignment in BML software.}
\label{lis:MA-ON}
\end{figure}
Figure~\ref{fig:MA-All} is comparison of the run-time of the benchmark (BHMK) vs. optimized (TUNED) micro-kernels for 
multiple dual socket Intel architectures using Intel 17 compiler. The performance of the optimized (TUNED) micro-kernel 
is at best 11X for high thread counts and 5X on average across all architectures.
%
\subsection{Micro-Kernel Performance Evaluation - Thread Binding and Affinity:}
\label{section:AB}
The purpose of rectifying the previously utilized thread binding and affinity settings used while performing 
linear algebra computations in BML~\cite{bml} is to ensure that sub-optimal hardware utilization is prevented. Two 
known issues that degrade performance when using multi-core computers are 1.) thread migration and 2.) the 
utilization of non-equivalent threads during simulation. Thread migration is a process whereby operational 
threads migrate or move from core to core during simulations. This behaviour affects the repeatably during 
performance testing or bench-marking. Non-equivalent thread utilization is a scenario whereby operational 
threads do not have the same associated resources (L1, L2 or L3/LLC cache). This also affects repeatably 
during performance testing or bench-marking. It occurs as a result of a lack of specificity in the environment 
variables that control the binding and affinity of thread to hardware core(s) and/or socket(s).  
Preventing thread migration and appropriately specifying the correct thread binding and 
affinity using Intels's API~\cite{VictorEijkhout-1,eijkhout2017parallel,Intel-2019-1,Intel-2019-2}, requires 
setting two environment variables namely $\texttt{KMP\char`_ AFFINITY}$ and $\texttt{KMP\char`_HW\char`_SUBSET}$. 
$\texttt{KMP\char`_ AFFINITY}$ controls the thread placement which is highly dependent on the predominant 
arithmetic intensity (AI) of the application. Figure~\ref{fig:NUMA__unaware} is a pictorial representation of 
a thread placement specified as ``scatter". In addition, setting both environment variable prevent thread 
migration. $\texttt{KMP\char`_HW\char`_SUBSET}$ determines the number of active threads and is set in the 
following format $\texttt{<\#1>s,2t,<\#2>c}$. \textup{c},\textup{t},\textup{s} stand for cores per socket, 
threads per core, and number of active sockets, respectively. As an example, for a dual socket hardware with 
twenty-four cores and two hardware threads per core, $\texttt{KMP\char`_HW\char`_SUBSET}$ set to 
$\texttt{1t,2s,24c}$ implies that forty-eight threads are operational on that node at one thread per core. 
Similarly, $\texttt{2s,2t,24c}$ implies that forty-eight threads are operational on each socket with a total of 
ninety-six total threads (two threads per core). Table~\ref{aff-and-bind-setting} shows the best practice 
guidelines for configuring the affinity and binding of an application depending on the procedure with the 
dominant AI using Intel and OpenMP API.
%
\input{tables/affinity}
\begin{figure}
\includegraphics[width=\textwidth]{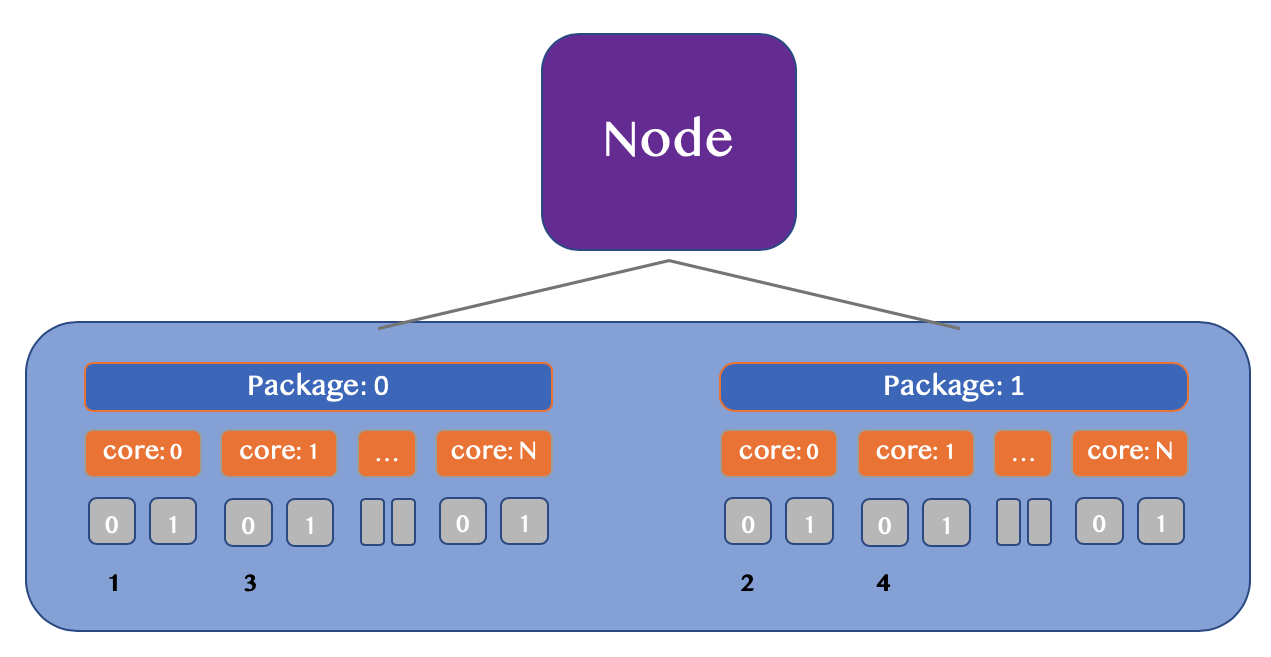}
\caption{An illustration of a modern dual socket multi-core architecture. 
             Showing a coordination of memory initialization with thread binding 
             and affinity with considerations for “First Touch” policies on 
             Linux systems.}
\label{fig:NUMA__unaware}
\end{figure}
%

%% file: tables/arch.tex

\begin{table}[h!]
\small
\begin{center}
\begin{tabular}{ |p{4cm}||p{2.5cm}|p{3cm}|p{2.5cm}|p{2cm}| }
\hline
\multicolumn{5}{|c|}{\text{Hardware Specification}} \\
\hline
Features:                              & Skylake-Gold & Skylake-Platinum & Cascade Lake & Cascade Lake w/ Optane \\
\hline
\text{Model Number}                    & 6152    & 8176     &  6254   &  8260  \\
\text{Launch Date}                     & Q3'17   & Q3'17    &  Q2'19  &  Q2'19 \\
Cores                                  & 22      & 28       &  18     &  24    \\
Threads Per Core                       & 2       & 2        &  2      &  2     \\
\text{Base Frequency(GHz)}             & 2.10    & 2.10     &  3.10   &  2.40  \\
\text{Turbo Frequency(GHz)}            & 3.70    & 3.80     &  4.10   &  3.90  \\
\text{Cache L3(MB)}                    & 30.25   & 38.50    &  24.75  &  35.75 \\
\text{HBM (MB)}                        & No      & No       &  No     &  No    \\
TDP (W)                                & 140     & 165      &  200    &  165   \\
\hline
\end{tabular}
\caption{ Hardware specification for Intel architectures (Ark Intel). }
\label{tab:hardware__specs}
\end{center}
\end{table}

%% file: tables/affinity.tex
\begin{table}[]
\centering
\begin{tabular}{ |p{5cm}||p{4cm}|p{4cm}|  }
\hline
\multicolumn{3}{|c|}{Affinity and Bindings Settings} \\
\hline
\multicolumn{3}{|l|}{Intel's API:} \\
\hline
\textbf{Arithmetic-Intensity}&\texttt{KMP\char`_HW\char`_SUBSET}  & \texttt{KMP\char`_ AFFINITY}\\
\hline
Compute Bound   & All HW Threads              & ``compact"    \\
Memory Bound    & $\frac{1}{2}$ HW Threads    & ``scatter"    \\
\hline
\multicolumn{3}{|l|}{Migration Control: set \texttt{KMP\char`_AFFINITY} \& \texttt{KMP\char`_HW\char`_SUBSET} } \\
\hline
\hline
\multicolumn{3}{|l|}{OpenMP's API:} \\
\hline
Compute Bound   & All HW Threads              & close    \\
Memory Bound    & $\frac{1}{2}$ HW Threads    & spread    \\
\hline
\multicolumn{3}{|l|}{Migration Control: set \texttt{OMP\char`_PROC\char`_BIND=true}, 
\texttt{OMP\char`_NUM\char`_THREADS} \& \texttt{OMP\char`_PLACES=cores}
} \\
\hline
\end{tabular}
\caption{Best practice guidelines for configuring the affinity and binding setting depending on predominant 
arithmetic intensity using Intel's~\cite{VictorEijkhout-1,eijkhout2017parallel,Intel-2019-1,Intel-2019-2} and OpenMP's~\cite{VictorEijkhout-1,eijkhout2017parallel} API.}
\label{aff-and-bind-setting}
\end{table}

%% file: results.tex
\section{Results}
\label{section:results}
\subsection{Brief}
\label{section:results-brief}
In order to demonstrate of the viability of the aforementioned optimizations we proceed following the three steps 
delineated below. First, we generate chemically relevant system matrices following the techniques discussed in 
section~\ref{section:results-1} representing Metals, Semi Conductors and Soft Matter. Second, we evaluate the 
performance of BML ELLPACK matrix-matrix multiply algorithm for each system. Third, in order to demonstrate the 
viability of the aforementioned optimizations in a full simulation, we use ExaSP2, a spectral 
projection proxy application\cite{SMniszewski-1} for carrying out QMD computations. Following that, we establish 
a benchmark run-time for ExaSP2 to determine signature of the SP2-Basic algorithm which will inform on the 
condition in which the best performance gains in run-time is achieved. After establishing the 
representative/dominant AI, we compare the performance of the BHMK to that of the TUNED for those specific cases. 
%
\subsection{On the Generation of Model Hamiltonian System Matrices}
\label{section:results-1}
\input{figures/model-Hamiltonian}
Chemically relevant system matrices typically fall under the following categories namely 1.) Metals, 
2.) Semi Conductors or 3.) Soft Matter. Figure~\ref{fig:modelH} is a schematic representation of the model 
Hamiltonian matrices. These model systems are constructed by coupling arrays of two-level systems with A and 
B as atomic type of orbitals. Given that, A and B orbitals have onsite energies $\epsilon_{\textup{A}}$ and 
$\epsilon_{\textup{B}}$, respectively. A coupling between the same type of orbital, that is, comprising of only 
A type or B type, is given by either $\delta_{\textup{A}_{i}\textup{A}_{j}}$ or $\delta_{\textup{B}_{i}\textup{B}_{j}}$. 
Similarly, a coupling of elements between different orbitals, that is A and B, is given by 
$\delta_{\textup{A}_{i}\textup{B}_{j}}$. All couplings between orbitals are modulated by an exponential dumping 
factor computed as $\exp(\textup{k}|\textup{j}-\textup{i}|)$, where k is the decaying constant, and i and j 
are the positions of both orbitals. All the couplings and onsite energies are in units of electron Volts(eV). 
We also introduced a randomization parameter that adds noise to couplings and onsite energies. This noise is 
introduced as $\textup{param}(1 + r \times \textup{RAND})$, where RAND is a random number chosen between -1 and 1 
and param represents any of the coupling or onsite energies involved.
A module that enables the generation of 
these Hamiltonian system matrix has been recently added to the PROGRESS~\cite{2016progress} library. 
Figure~\ref{fig:systemsH} is a plot of the total DOS computed out of different model Hamiltonian matrices. The 
Fermi Level of the system is set to be 0.0 eV. System matrices representing Metals were generated by setting:
\begin{equation*}
\delta_{A_iA_j} = -1.0, \;\;\; \delta_{B_iB_j} = -1, \;\;\; k = -0.01,
\end{equation*}
and the rest of the parameters to 0.0, resulting in a sparsity of 0.98\%. For system matrices representing Semi 
Conductors, the parameters set to:
\begin{equation*} 
\delta_{B_iB_j} = -1.0, \;\;\; \delta_{AB} = -2.0, \;\;\; k = -0.01,
\end{equation*}
and the rest of the parameters to 0.0, resulting in a sparsity of 94\% sparse. Soft Matter systems matrices were 
generated by setting:
\begin{equation*}
\delta_{B_iB_j} = -1.0, \;\;\; \delta_{AB} = -1.0, \;\;\; \epsilon_A = -10.0, \;\;\; k = -0.1, \;\;\; r = 1.0,
\end{equation*}
resulting in a sparsity of 82\% sparse. Figure~\ref{fig:MatrixPlot-Metals-SemiCond-SoftMatt} is a matrix plot using 
the BML Ellpack format for Metals, Semi Conductors and Soft Matter.
\input{figures/totalDOS}
%
\subsection{Performance Evaluation using System Matrices}
\label{section:results-2}
\begin{figure}
\includegraphics[width=\textwidth]{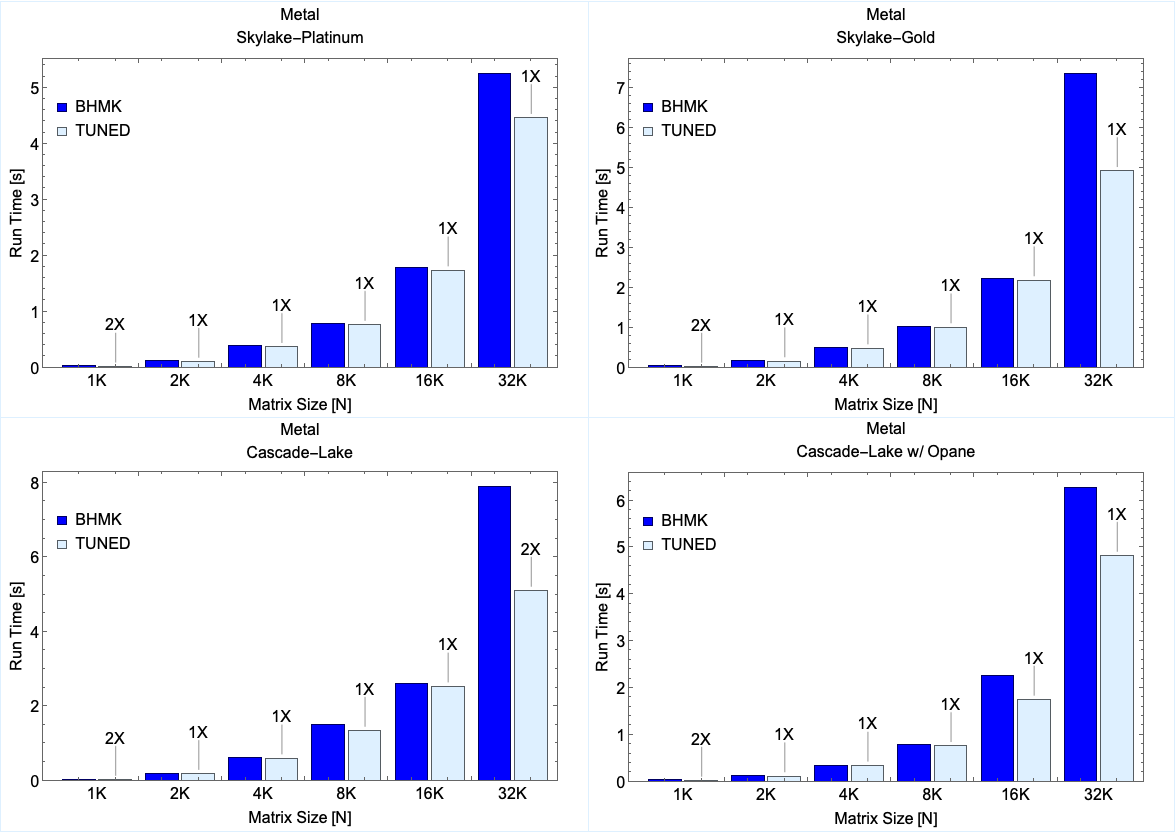}
\caption{Performance in run-time for the benchmark (BHMK) vs. optimized (TUNED) BML ELLPACK matrix-matrix multiply for 
         Metals computed on Cascade Lake (Top-Left), Sky-Lake Gold (Top-Right), 
         Sky-Lake Platinum (Bottom-Left) and Cascade Lake with Intel Optane Technology (Bottom-Right).}
\label{fig:BMLEllpackX2-Metals}
\end{figure}

\begin{figure}
\includegraphics[width=\textwidth]{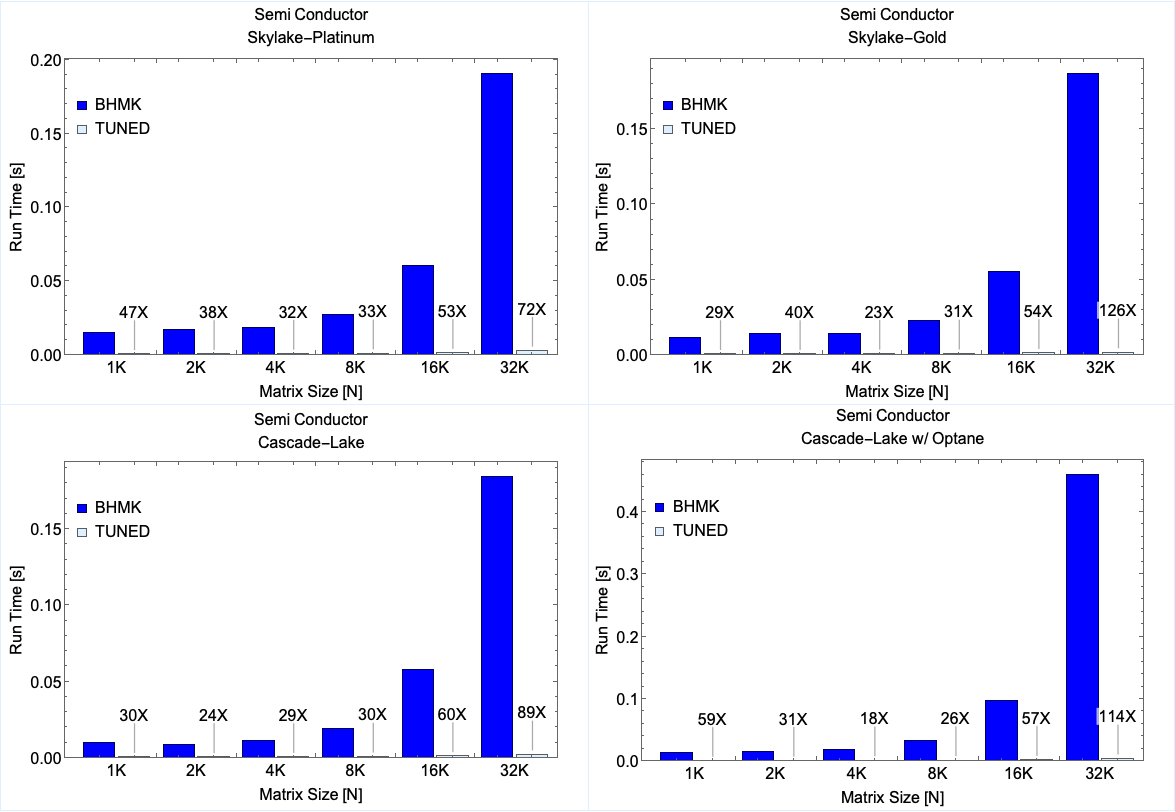}
\caption{Performance in run-time for the benchmark (BHMK) vs. optimized (TUNED) BML EllPACK matrix-matrix multiply 
         for Semi Conductor computed on Cascade Lake (Top-Left), Sky-Lake Gold (Top-Right), 
         Sky-Lake Platinum (Bottom-Left) and Cascade Lake with Intel Optane Technology (Bottom-Right).}
\label{fig:BMLEllpackX2-SemiConductor}
\end{figure}

\begin{figure}
\includegraphics[width=\textwidth]{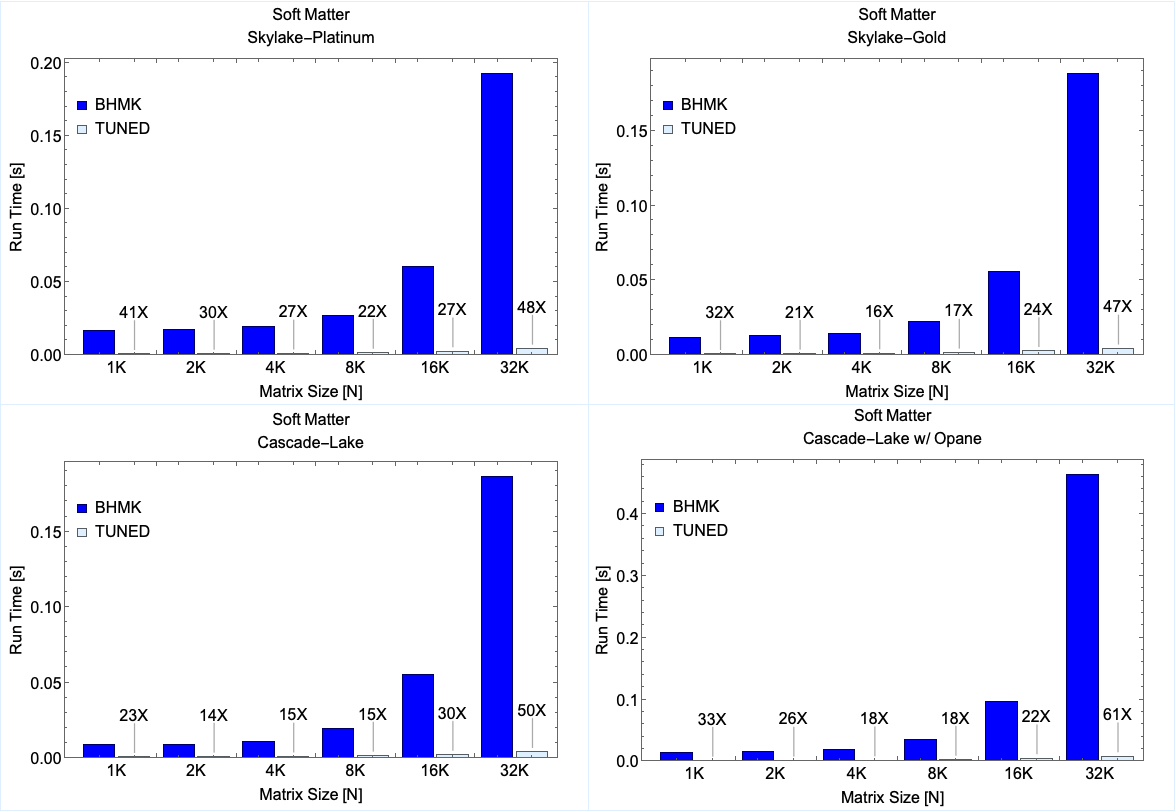}
\caption{Performance in run-time for the benchmark (BHMK) vs. optimized (TUNED) BML EllPACK matrix-matrix multiply 
         for Soft Matter computed on Cascade Lake (Top-Left), Sky-Lake Gold (Top-Right), 
         Sky-Lake Platinum (Bottom-Left) and Cascade Lake with Intel Optane Technology (Bottom-Right).}
\label{fig:BMLEllpackX2-SoftMatter}
\end{figure}

Figure~\ref{fig:MatrixPlot-Metals-SemiCond-SoftMatt} is a matrix plot of three systems namely metals, semi conductors 
and soft matter. It serves as a visual representation of the density of each system; with metal being the most dense. 
Figure~\ref{fig:BMLEllpackX2-Metals}, \ref{fig:BMLEllpackX2-SemiConductor} and \ref{fig:BMLEllpackX2-SoftMatter} show 
the performance of the benchmark (BHMK) vs. optimized (TUNED) BML matrix-matrix multiply 
algorithm using the ELLPACK format for the aforementioned systems on four different architectures. Matrix-matrix 
multiplications with metals (Figure~\ref{fig:BMLEllpackX2-Metals}), shows a 15\% to 30\% improvement on average in the 
optimized version of the BML library for larger matrix sizes on every architecture. Semi conductors 
(Figure~\ref{fig:BMLEllpackX2-SemiConductor}), show excellent improvement in the optimized code for all matrix sizes 
experimented with. On average, the 32k matrices are improved by over 100 folds. Figure~\ref{fig:BMLEllpackX2-SoftMatter}, 
(soft matter systems) shows similar improvements to that semi conductors. On average, the 32k matrices are improved by 
over 50 folds across all architectures.
\input{figures/matrix-plot-1}
%
\subsection{Performance Evaluation using an ExaSP2 Application:}
\label{section:results-3}
\begin{figure}
\includegraphics[width=\textwidth]{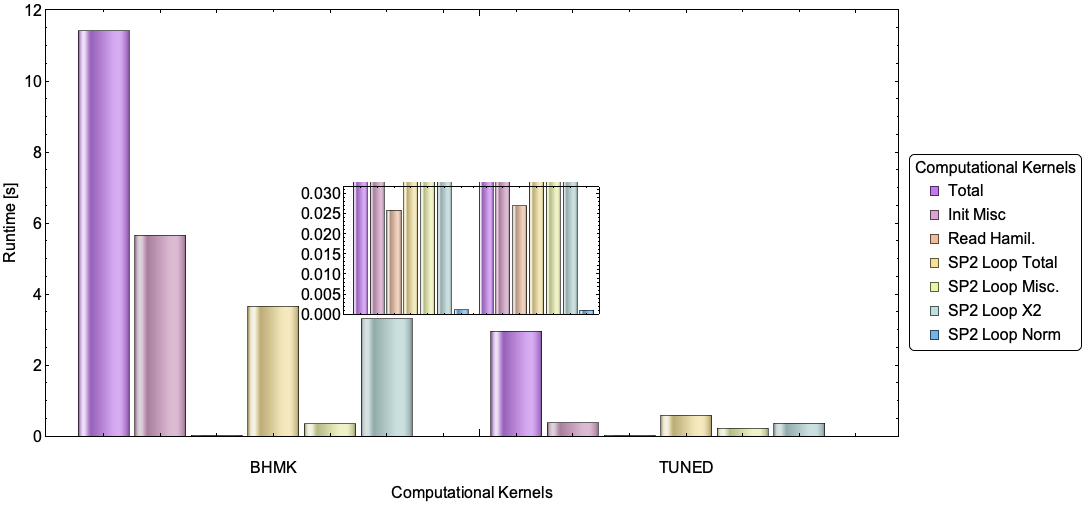}
\caption{Performance in run-time for the benchmark (BHMK) vs. optimized (TUNED) ExaSP2 for 
         Semi Conductor simulated on Intel's Cascade Lake. Showing run-time of individual subroutines.}
\label{fig:ExaSP2SemiCondCascadeLake}
\end{figure}

\begin{figure}
\includegraphics[width=\textwidth]{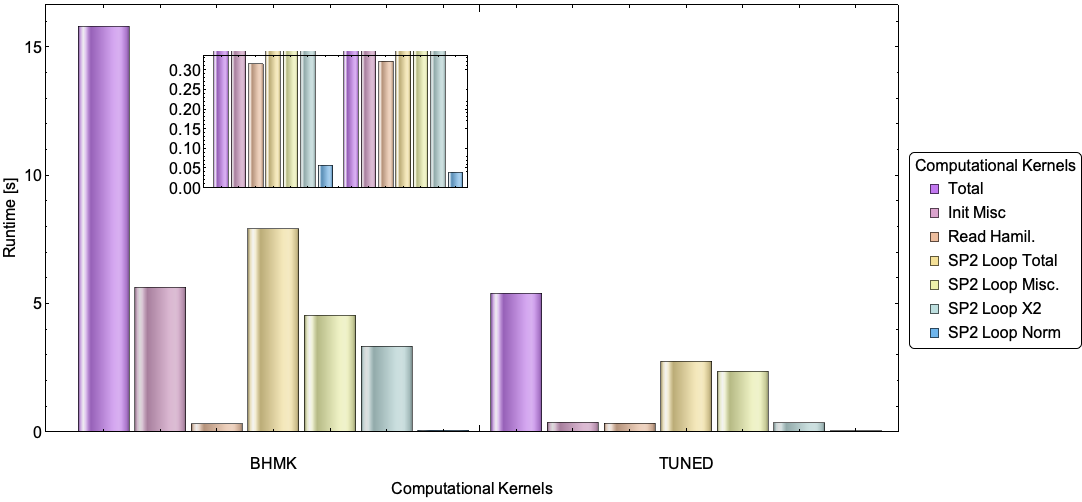}
\caption{Performance in run-time for the benchmark (BHMK) vs. optimized (TUNED) ExaSP2 for 
          Soft Matter simulated on Intel's Cascade Lake. Showing run-time of individual subroutines.}
\label{fig:ExaSP2SoftMattCascadeLake}
\end{figure}

\begin{figure}
\includegraphics[width=\textwidth]{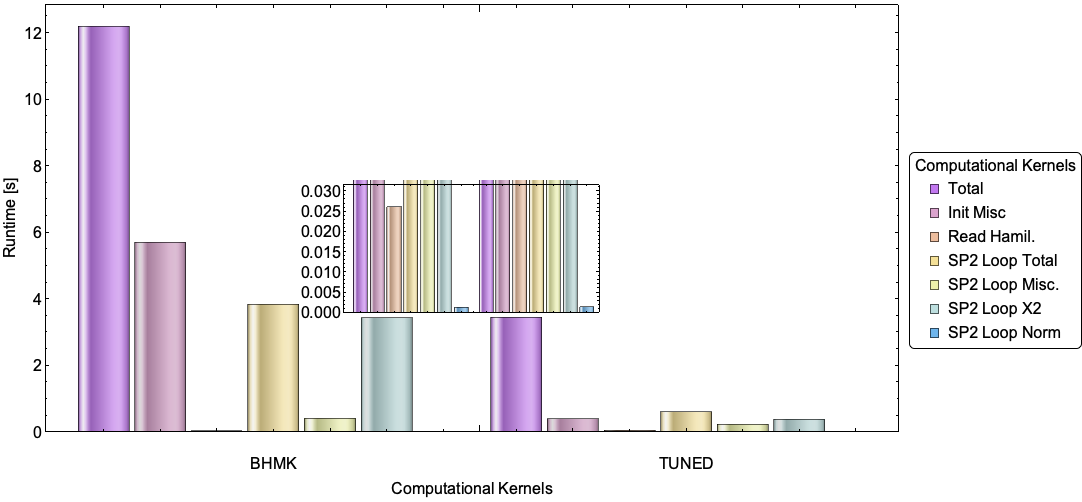}
\caption{Performance in run-time for the benchmark (BHMK) vs. optimized (TUNED) ExaSP2 for 
         Semi Conductor simulated on Intel's Skylake Platinum. Showing run-time of individual subroutines.}
\label{fig:ExaSP2SemiCondSkylakePlatinum}
\end{figure}

\begin{figure}
\includegraphics[width=\textwidth]{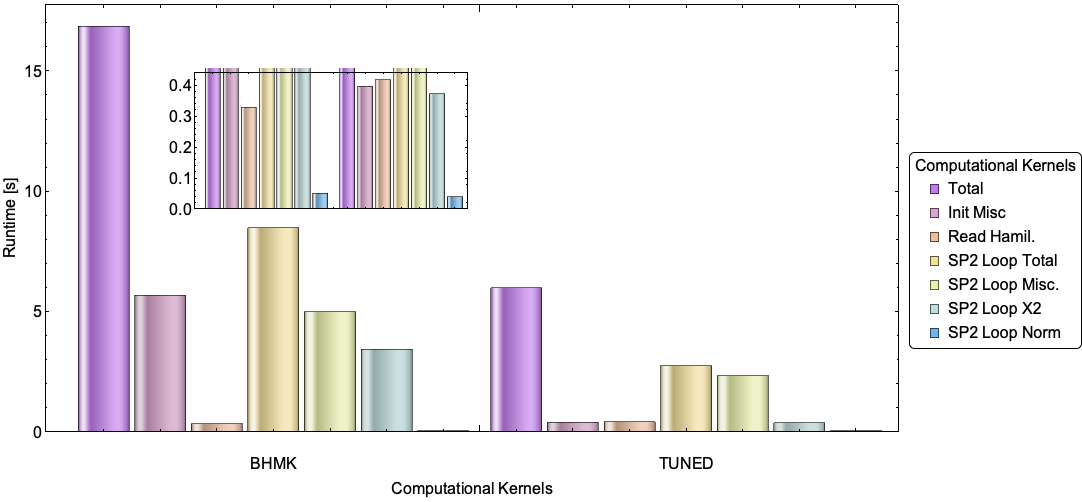}
\caption{Performance in run-time for the benchmark (BHMK) vs. optimized (TUNED) ExaSP2 for 
          Soft Matter simulated on Intel's Skylake Platinum. Showing run-time of individual subroutines.}
\label{fig:ExaSP2SoftMattSkylakePlatinum}
\end{figure}

The procedures that make up the SP2-Basic algorithm can be decomposed into two types namely the initialization 
and SP2 calculation step. The initialization step involves reading in the Hamiltonian (Read Hamiltonian) matrix 
and other miscellaneous initialization sub-steps (Init. Misc.). The SP2 calculation step involves a matrix-matrix 
multiplication (SP2 Loop X2) and a matrix norm (SP2 Loop Norm) calculation. All other steps within the SP2 
calculation are categories under (SP2 Loop Misc.). For this exercise, only Semi Conductor and Soft Matter where
experimented with as they are appropriate for an SP2 calculation. The ELLPACK matrix format and corresponding 
algorithms were used to represent both systems. In addition, since the overall performance of each system has 
shown repeat-ability across all four architectures, only Intel Cascade Lake and Skylake Platinum were used here. 
Figure~\ref{fig:ExaSP2SemiCondCascadeLake}, \ref{fig:ExaSP2SoftMattCascadeLake}, 
\ref{fig:ExaSP2SemiCondSkylakePlatinum} and \ref{fig:ExaSP2SoftMattSkylakePlatinum} is a comparison of the 
benchmark (BHMK) vs. optimized (TUNED) ExaSP2 for both systems. On Intel Cascade Lake 
(figure~\ref{fig:ExaSP2SemiCondCascadeLake} and \ref{fig:ExaSP2SoftMattCascadeLake}), both systems show an 
improvement in run-time by over four folds (4x) while other subroutines vary from four (4x) to twelve (12x) folds.
%

%% file: figures/model-Hamiltonian.tex
\begin{figure}
    \centering
    \begin{tikzpicture}[->,>=stealth,shorten >=1pt,auto,node distance=4cm,
                thick,main node/.style={circle,draw,font=\Large\bfseries}]
  \node[main node] (b) {B$_i$};
  \node[main node] (a) [below of=b] {A$_i$};
  \node[main node] (b2) [right of=b] {B$_j$};
   \node[main node] (a2) [below of=b2] {A$_j$};
   \node[main node] (p1) [left of=b] {};
   \node[main node] (p2) [right of=b2] {};
   \node[main node] (p3) [left of=a] {};
   \node[main node] (p4) [right of=a2] {};

  \path[-]
    (a) edge node [dashed] {$\delta_{AB}$} (b)
    (a2) edge node {$\delta_{AB}$} (b2)
    (b) edge node {$\delta_{B_iB_j}$} (b2)
    (a) edge node {$\delta_{A_iA_j}$} (a2);
  \path[dashed]
    (b) edge (p1)
    (b2) edge (p2)
    (a) edge (p3)
    (a2) edge (p4)
    ;
\end{tikzpicture}
    \caption{Two-level system model Hamiltonian used to generate representative Hamiltonian matrices for bench-marking purposes. The model has the following parameters: Four coupling parameters, four onsite energies, a decaying exponential parameter, and a randomization factor. }
    \label{fig:modelH}
\end{figure}{}

%% file: figures/totalDOS.tex
\begin{figure}
    \centering
    \includegraphics[scale=0.18]{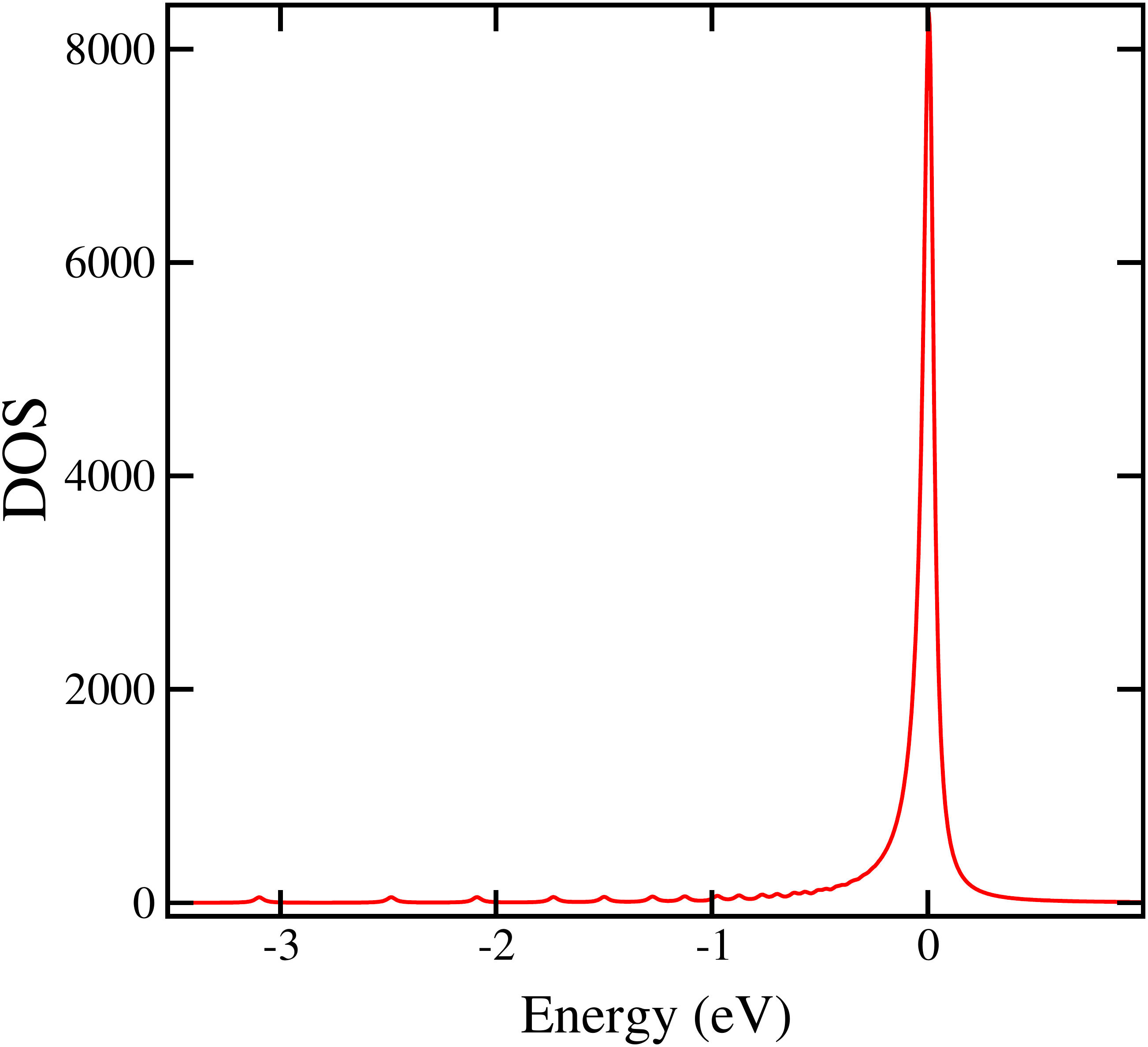}
    \quad
    \includegraphics[scale=0.18]{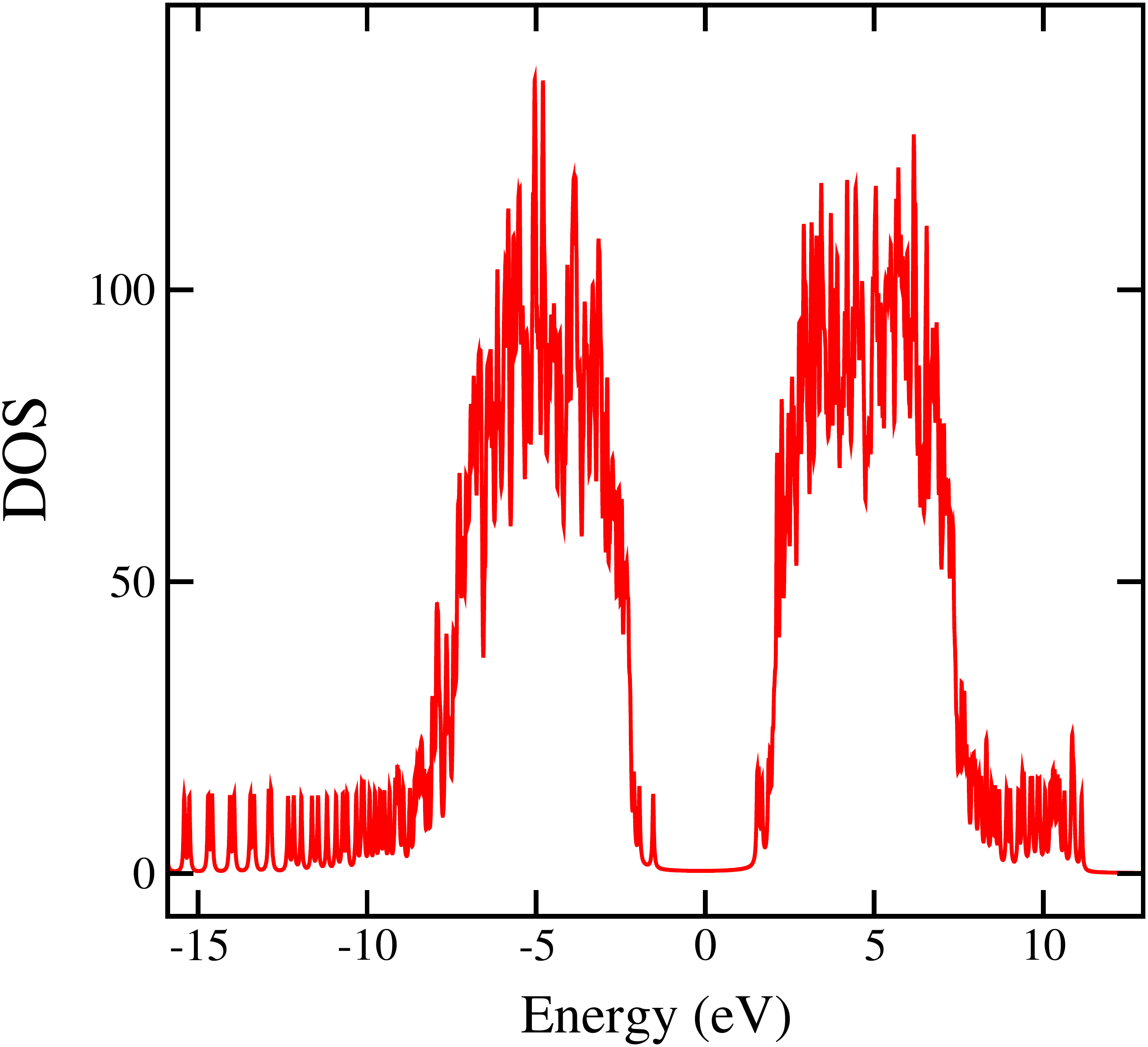}
    \quad
    \includegraphics[scale=0.18]{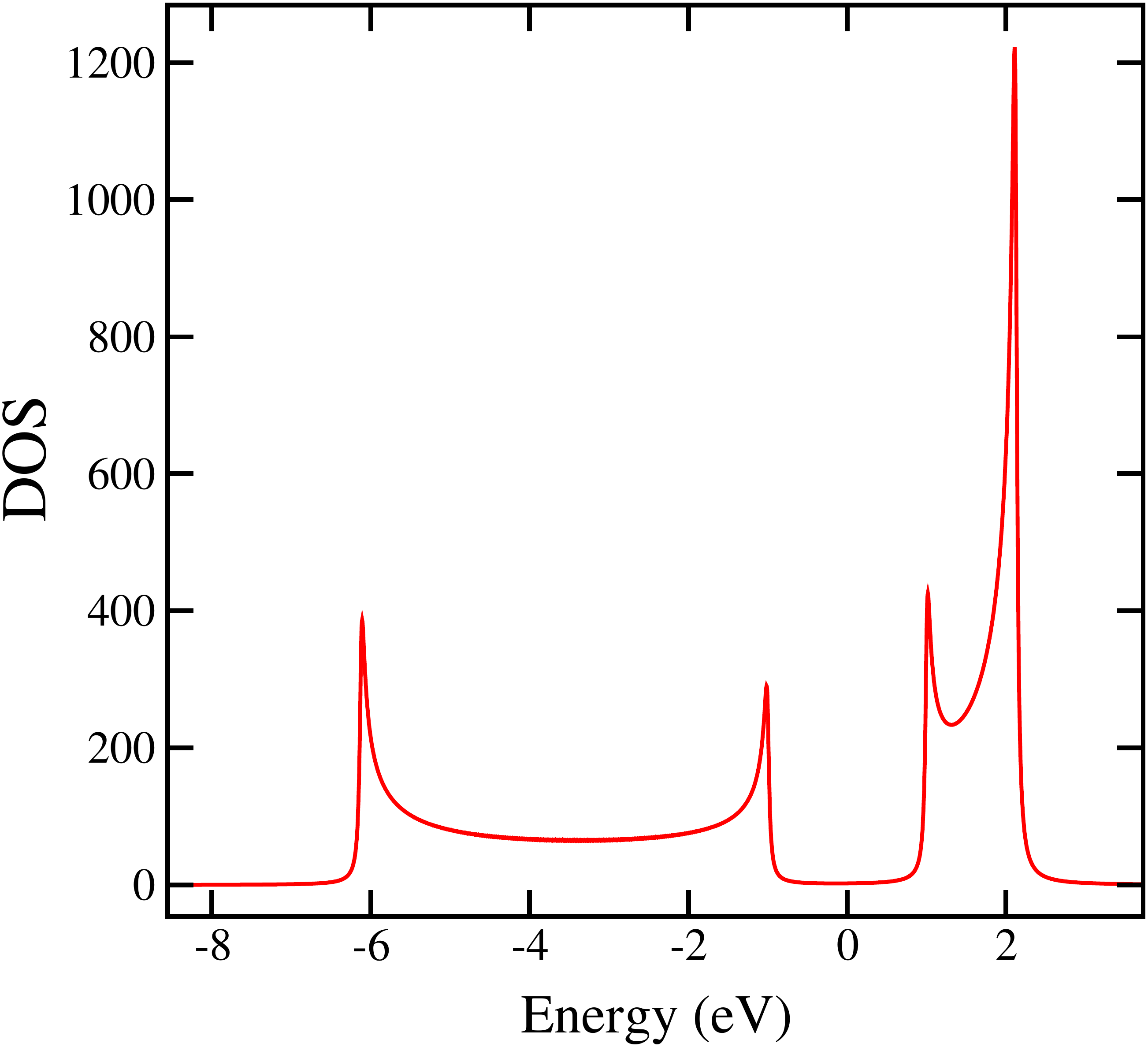}
    \caption{Total DOS computed out of different model Hamiltonian matrices. The Fermi Level of the system is set to be 0.0 eV. a) Metals can be generated by setting $\delta_{A_iA_j} = -1.0$, $\delta_{B_iB_j} = -1$, $k = -0.01$, and the rest of the parameters to 0.0. Matrices of this type are about 0.98\% sparse. b) Semiconductors can be generated by setting $\delta_{B_iB_j} = -1.0$ and $\delta_{AB} = -2.0$, $k = -0.01$ and the rest of the parameters to 0.0. Matrices of this type are about 94\% sparse. c) Soft matter systems can be generated by setting $\delta_{B_iB_j} = -1.0$ and $\delta_{AB} = -1.0$, $\epsilon_A = -10.0$, $k = -0.1$ and $r = 1.0$. Matrices of this type are about 82\% sparse.}
    \label{fig:systemsH}
\end{figure}{}

%% file: figures/matrix-plot-1.tex
\begin{figure}
\centering
\subfloat[Metals]{\includegraphics[width=65mm]{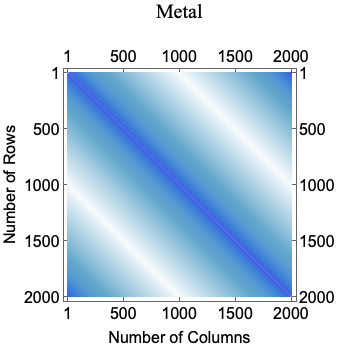}}
\subfloat[Semi Conductors]{\includegraphics[width=65mm]{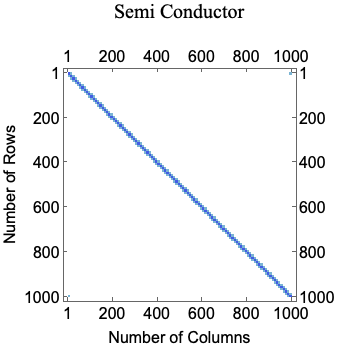}}
\hspace{0mm}
\subfloat[Soft Matter]{\includegraphics[width=65mm]{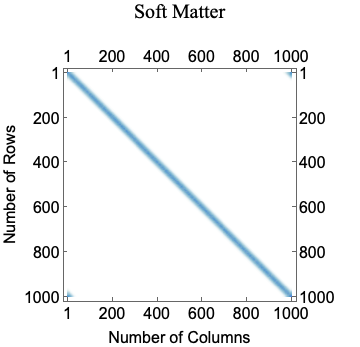}}
\caption{Matrix plot of Metals (Top Left), Semi Conductors (Top Right) and Soft Matter 
         (Bottom) generated using the techniques discussed.}
\label{fig:MatrixPlot-Metals-SemiCond-SoftMatt}
\end{figure}

%% file: summary.tex
\section{Conclusions}
\label{section:conclusion}
We evaluated the viability of several optimization strategies namely 1.) Strength Reduction 2.) Memory 
Alignment 3.) NUMA aware allocations and all incorporation with the appropriate thread affinity and binding. 
Initially, we tested the viability of these performance tuning techniques in micro-kernels by comparing the 
run-time before and after optimization for several target architectures. For SR optimizations, the optimized 
micro-kernel showed approximately 2X across multiple thread sizes and varying architectures. For NUMA aware 
data initialization that ensures data locality, the performance of the optimized micro-kernel is at best 3X for 
high thread counts and 2X on average on all architectures. Memory alignment optimizations, the optimized 
micro-kernel is at best 11X for high thread counts and 5X on average across all architectures. The optimizations 
where used in conjunction with an appropriate thread binding and affinity settings such that, two well known 
issues that degrade performance when using multi-core computers are avoided. 

We followed this up by generating chemically relevant system matrices representing Metals, Semi Conductors and 
Soft Matter evaluate the performance of BML ELLPACK matrix-matrix multiply algorithm for each system. 
The matrix-matrix multiplications with metals showed a 15\% to 30\% improvement on average in the 
optimized version of the BML library for larger matrix sizes on every architecture. Semi conductors, 
showed an excellent improvement in the optimized code for all matrix sizes experimented with. On average, the 
32k matrices were improved by over 100 folds. Soft matter systems showed similar improvements to that semi conductors
where on average, the 32k matrices are improved by over 50 folds across all architectures. Finally, we evaluated 
the performance of these optimizations using a QMD proxy application, where the ELLPACK matrix format and 
corresponding algorithms were used to represent both Soft Matter and Semiconductors. Both systems on Intel Cascade 
Lake and Skylake Platinum showed an improvement in run-time by over four folds (4x). Other relevant subroutines 
within ExaSP2 showed improvements varying from four (4x) to twelve (12x) folds.

\normalfont
\clearpage